\documentclass[conference]{IEEEtran}
\IEEEoverridecommandlockouts
\usepackage{cite}
\usepackage{amsmath,amssymb,amsfonts}
\usepackage{algorithmic}
\usepackage{graphicx}
\usepackage{ragged2e}
\usepackage{textcomp}
\usepackage{xcolor}
\def\BibTeX{{\rm B\kern-.05em{\sc i\kern-.025em b}\kern-.08em
    T\kern-.1667em\lower.7ex\hbox{E}\kern-.125emX}}
\begin{document}

\title{Segmentation and Shape Detection Techniques In Ultrasound Images - A Review\\
}

\author{\IEEEauthorblockN{1\textsuperscript{1st} Subho Ghose}
\IEEEauthorblockA{\textit{dept. Computing Science} \\
\textit{University of Alberta}\\
Edmonton, Canada \\
subho@ualberta.ca}
\and
\IEEEauthorblockN{2\textsuperscript{nd} Haixia Wu}
\IEEEauthorblockA{\textit{dept. Computing Science} \\
\textit{University of Alberta}\\
Edmonton, Canada \\
haixia2@ualberta.ca}
\and
\IEEEauthorblockN{3\textsuperscript{rd} Ruturaj Gole}
\IEEEauthorblockA{\textit{dept. Computing Science} \\
\textit{University of Alberta}\\
Edmonton, Canada \\
ruturaj@ualberta.ca}
%Add any more group member or author who contributed to this work
}

\maketitle

\begin{abstract}
Medical ultrasound imaging can prove paramount in detection of unseen abnormalities such as tumors and deformed organs. Regardless of their importance, the interpretation of these images is highly dependent on varying subjective experts like radiologists. Automation is therefore required to objectively standardize the process of segmentation and detection of regions in Ultrasound images. However, this task isn't a straightforward one as issues exist such as speckle noise, that can take away important features if removed, or the excess of features that amount of multiple false positives. We present in this paper a review of the many approaches that have been taken to deal with these obstacles and get a better accuracy at segmentation and detection of regions. As different organs and regions pose different challenges, the paper has been divided by the respective clinical application related to the work. A list of generic approaches has also been detailed in order to not rule out the possibility of extending these works to more specialized tasks. A plethora of machine and deep learning techniques have been implemented since as far back as the 1990's. Their performances and efficiencies have been compared and commented upon. The proposed research of ours also aims to first start with simple techniques like derivating the image to detect edges, and then to compare deep learning methods such as Dual Path Networks with machine learning techniques.
\end{abstract}

\section{Introduction}
\label{intro}
Ultrasound is a technique widely used for various applications such as diagnosis of diseases or simple studies of organs in medical settings, as well as applications related to fields as diverse as weapons, wireless communication, medicine, therapy [1] and even TV remotes [2]. The simple, straightforward functioning of ultrasonic scanners, that is, the emission and receival of ultrasonic waves to map surfaces, makes them very popular in understanding the human body. Moreover, this has led to revolutionary advancements such as the detection and identification of abnormalities and anomalies in the body like cancerous tumors. However, despite their usefulness, currently they mostly offer good results only with the assistance of an expert, namely a radiologist. Subjective opinions can differ from radiologist to radiologist [3], not to mention the prevalence and possibility of human error that isn't easily correctable. Thus, not only are objective standards necessary in order to understand and analyze an Ultrasound image, automation is necessary to enforce consistent standardized assessments. Medical ultrasound imaging can come in many flavours, from 2D Brightness Mode (B-Mode), Amplitude Mode (A-Mode), Doppler Mode to 3D and even 4D scanners that can map the surfaces in 3 dimensions with an additional 4th dimension of time [4]. Unfortunately, segmentation of Ultrasound images is not a straightforward task as there are many artefacts that need to be dealt with. Principally, these images are not the most clear images, and it is difficult to separate the noise from important features. The speckle noise that is most prevalent in these images can sometimes be a feature itself that shouldn't be discarded [5]. Removal of noise might as well result in the worsening of the detection. Even if noise removal was warranted, there isn't a good understanding about the origin of the noise in these images, which is why it becomes more difficult to eliminate it [6]. This allows for the possibility of many false positives. Other than these surface issues, orientation is a problem that needs to be considered [7]. The subject in question needs to be scanned from different orientations for a better imaging. Blood flow and other such obstructions might cause distortions, occlusions and contribute to noise. Physical characteristics of the subject being scanned should be accounted for, as these variations would manifest in the image differently for different people. Moreover, the organs and parts of the body themselves are physiologically and anatomically so different that one generic method cannot be guaranteed to work with all. Limited datasets pose a problem, as although there are open-access datasets available, one has to rely on them without first-hand getting the scans if there isn't an access to the scanner. The presented literature review aims to evaluate different methods of segmentation to get a better idea for detection and recognition of shapes in Ultrasound images as well as 3D reconstruction of the same. Methods are classified according to their clinical applications and then there are some generic approaches listed which can be extended to shape detection in Ultrasound image. 

\section{Literature Review}
\label{LR}
The review has been divided into sections by the clinical applications related to the work. Later on, generic approaches have also been considered. \newline
\textbf{Clinical Applications:} \newline
\textbf{Breast} \newline
i)
In 2002, Drukker and his team investigated the use of a radial gradient index (RGI) filtering technique to automatically detect lesions on breast ultrasound. Here, they proposed a computer-aided diagnosis (CAD) method to improve lesion detection in breast, which is computationally tractable, allowing for future implementation in real-time sonography. In their experiment they used round robin analysis to assess the quality of the classification of lesion candidates into actual lesions and false-positives by a Bayesian neural network. Although the RGI-filtering does not use ultrasound specific information, such as shadowing; however, it is sometimes confused by such ultrasound image characteristics. So, they claimed combining the RGI-filtering technique with ultrasound specific detection methods as a promising approach to improve the detection rate of solid masses and reduce the number of false-positive detections [8].\newline
ii) 
In 2002, Giger and his team proposed a method for breast lesions on ultrasound named Computer-aided diagnosis (CAD) that is based on the automatic segmentation of lesions and the automatic extraction of four features related to the lesion shape, margin, texture, and posterior acoustic behavior. It uses clinically motivated, computer-extracted sonographic features to quantify the four features. The computer-extracted features are then merged through linear discriminant analysis. Three studies were performed on a large clinical database of 400 cases: (1) evaluation of the marginal benefit of each feature to our CAD method; (2) determination of the performance of our CAD method in distinguishing carcinomas from different types of benign lesions; and (3) independent validation of the method using 11 independent trials. The main limitation of the proposed method is that they failed to show a statistically significant difference between the best two feature classifier and the four-feature classifier [9].\newline
iii) Computerized tumor segmentation on breast ultrasound (BUS) images
remains a challenging task. In this paper, we proposed a new method
for semi-automatic tumor segmentation on BUS images using Gaussian
filtering, histogram equalization, mean shift, and graph cuts [10].\newline
iv) We formulate the tumor detection as a two-step learning problem: tumor
localization by bounding box and exact boundary delineation. Specifically, the
proposed method uses an AdaBoost classifier on Harr-like features to detect a
preliminary set of tumor regions. The preliminarily detected tumor regions are
further screened with a support vector machine using quantized intensity features.
Finally, the random walk segmentation algorithm is performed on the US image to
retrieve the boundary of each detected tumor region [11].

\textbf{Kidney} \newline
i) Prevost and his team presented an automatic method to segment the kidney in 3D contrast-enhanced ultrasound (CEUS) images in 2012. The method is composed of two steps: first, the kidney is automatically localized by a novel robust ellipsoid detector; then, segmentation is obtained through the deformation of this ellipsoid with a model-based approach. To cope with low image quality and strong organ variability induced by pathologies, the algorithm allows the user to refine the result by real-time interactions. The proposed method is much faster, which requires only a few seconds. However, it can be challenging because of the noise, the artifacts and the partial occlusion of the organ (due to the limited field of view) [12].

ii) Marsousi et al, introduce a computer-aided kidney shape detection method suitable
for volumetric (3D) ultrasound images. Using shape and texture priors, the
proposed method automates the process of kidney detection, which is a problem
of great importance in computer-assisted trauma diagnosis. This paper
introduces a new complex-valued implicit shape model, which represents the
multiregional structure of the kidney shape. A spatially aligned neural network
classifiers with complex-valued output is designed to classify voxels into
background and multiregional structure of the kidney shape. The complex values
of the shape model and classification outputs are selected and incorporated in a
new similarity metric, such as the shape-to-volume registration process only fits
the shape model on the actual kidney shape in input ultrasound volumes. The
algorithm's accuracy and sensitivity are evaluated using both simulated and
actual 3-D ultrasound images, and it is compared against the performance of the
state of the art. The results support the claims about accuracy and robustness of
the proposed kidney detection method, and statistical analysis validates its
superiority over the state of the art [13].

iii) Xie et al, present a novel texture and shape priors based method for kidney segmentation in
ultrasound (US) images.Texture features are extracted by applying a bank of Gabor filters on
test images through a two-sided convolution strategy. The tex-ture model is constructed via
estimating the parameters of a set of mixtures of half-planed Gaussians using the expectation-
max-imization method. Through this texture model, the texture simi-larities of areas around
the segmenting curve are measured in the inside and outside regions, respectively. We also
present an itera-tive segmentation framework to combine the texture measures into the
parametric shape model proposed by Leventon and Faugeras.Segmentation is implemented by
calculating the parameters of the shape model to minimize a novel energy function. The goal
of this energy function is to partition the test image into two regions, the inside one with high
texture similarity and low texture variance,and the outside one with high texture variance.
The effectiveness of this method is demonstrated through experimental results on both natural
images and US data compared with other image segmentation methods and manual
segmentation [14].

\begin{figure}[htbp]
\centerline{\includegraphics[width=3in]{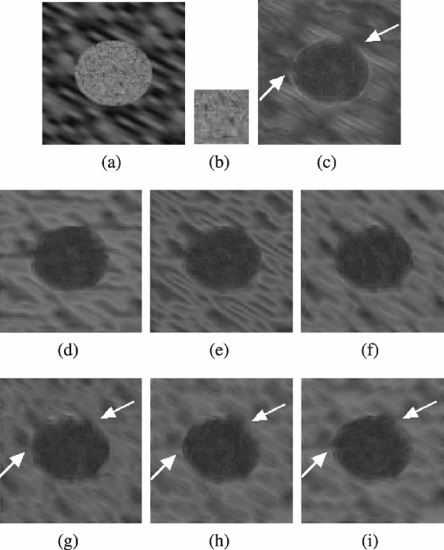}}
\caption{Fig. 1.
Texture similarity maps of a synthetic texture image. In those maps, high intensity indicates high difference. (a) the original synthetic texture image. (b) the training texture pattern. (c) the similarity map of the full-region convolution scheme. (d)-(i) the similarity maps of each upper half-plane for six different training directions. [14]}
\label{fig}
\end{figure}

\textbf{Fetal}\newline
i) To perform automatic extraction of standard planes for fetal anatomy assessment, 3D ultrasound (US) is a promising technique. However, this requires prior organ localization, which is difficult to obtain with direct learning approaches because of the high variability in fetus size and orientation in US volumes. So, in 2017,	Raynaud and his team proposed a methodology to overcome this spatial variability issue by scaling and automatically aligning volumes in a common 3D reference coordinate system. This preprocessing allows the organ detection algorithm to learn features that only encodes the anatomical variability while discarding the fetus pose. The proposed approach consists in tackling the successive difficulties step by step, by focusing on well identified problems to ensure the robustness of the whole processing pipeline. In order to increase the detection performance, directions such as the use of more volumes, data augmentation, image resolution refinement or separation of learning according to pregnancy stages are foreseen [15].

ii) In 2017, Gomez and his team proposed a fast feature-based rigid registration framework with a novel feature saliency detection technique. The method works by automatically classifying candidate image points as salient or non-salient using a support vector machine trained on points which have previously driven successful registrations. The resulting candidate salient points are used for symmetric matching based on local descriptor similarity and followed by RANSAC outlier rejection to obtain the final transform. After applying to data from 5 patients their method achieved similar accuracy and similar saliency detection quality in less than 5% the detection time, showing promising capabilities towards real-time whole-body fetal ultrasound imaging [16].

iii) In 2016, Sridar and his team introduced a general framework to automatically identify the different planes of any single fetal organ. They fine-tuned a pre-trained convolutional neural network (CNN) to create a feature extractor that derives the image features that are best for discriminating fetal ultrasound images without any reliance on anatomical priors or preprocessing. The method achieved a mean accuracy of 94.97% and 85.74% in the classification of fetal head and heart planes, which was higher than the state-of-the-art baseline algorithms. However, this achieved results with lower accuracy when there are so many variations in the images especially where the blood vessels appear to merge [17].

\textbf{Cardiology}\newline
i) Li et al., went further with the selection of Extremal Regions of Extremum Level (EREL). From a set of extremal regions, EREL detects the regions of interest. This problem comes from the segmentation of the borders of arterial walls in Intravascular (IVUS) images, like the one mentioned in the previous paper. After ordering EREL regions in an ascending order of the size of their features, the distances of their borders to their centers are measures and then they're put through 2D correlation coefficients to eliminate those regions that have lower correlation between them. An EREL is selected to find another EREL with the highest compactness measure to it. Although fast and straightforward, the model has poor performance for a small dataset of images and might choose a bad representation for classification of lumen [18].
 
ii) It is well known that segmentation of the left atrium and deriving its size can help to
predict and detect various cardiovascular conditions, as a result, automation of this
process in 3D Ultrasound image data is desirable to save costs and time in
echocardiographic laboratories.
Degel et al, introduce a combined deep-learning based approach on volumetric
segmentation in Ultrasound acquisitions with incorporation of prior knowledge about
left atrial shape and imaging device.
it is consists of three existing methods; 3D Fully Convolutional Segmentation
Network , Anatomic Constraint , and Domain Adaptation. Nevertheless, it is a novelty
to model the solution in a single framework, enabling analysis on the contribution of
each element on the primary segmentation task. Further, the domain adaptation
method has been leveraged to a 3D FCN segmentation framework, and applied
successfully to the Left atrium (LA), showing a statistical significant improvement

The results show, that including a shape prior helps the domain adaptation and the
accuracy of segmentation is further increased with adversarial learning [19].

iii) Yang et al, proposed a method based on the Fully Convolutional Network (FCN) to segment and delineate the lumen (interior walls) and media adventitia (exterior walls) in the IVUS (Intravascular Ultrasound) images. The method was named IVUS-Net, and its results were measured by Jaccard Measure (JM) and Hausdorff Distances (HD). The paper claims to be the first deep learning method used to segment the lumen and media adventitia. Data augmentation was required due to a limited dataset, and was done by flipping the ground truth images in different images as well as adding additive Gaussian noise. It is claimed that the method performs segmentation in 0.15 seconds and outperforms state-of-the-art methods [20].

\begin{figure}[htbp]
\centerline{\includegraphics[width=3in]{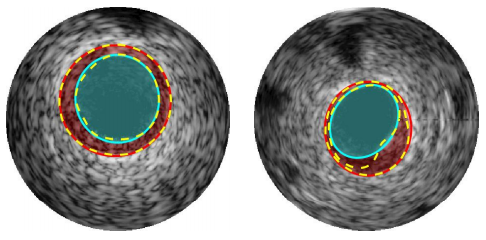}}
\caption{Fig.2: Lumen \& Media Adventitia Segmentation (Cyan \& Red respectively) and Gold Standard (Yellow) [20]}
\label{fig}
\end{figure}

iv) A Fully Convolutional Network (FCN) based deep architecture called
Dual Path U-Net (DPU-Net) is proposed for automatic segmentation
of the lumen and media-adventitia in IntraVascular UltraSound
(IVUS) frames.
 It introduces a deep architecture that is able to learn using a small
number of training images and still achieves a high degree of
generalization ability. Second, it strengthens the proposed DPU-Net
by having a real-time augmentor control the image augmentation
process [21]. 

v) The extraction of six standard planes in 3-D cardiac ultrasound plays an
important role in clinical examination to analyze cardiac function. A guideline-
based learning method for efficient and accurate standard plane extraction is
proposed. A cardiac ultrasound guideline determines appropriate operation steps
for clinical examinations. The idea of guideline-based learning is incorporating
machine learning approaches into each stage of the guideline. First, Hough forest
with hierarchical search is applied for 3-D feature point detection. Second, initial
planes are determined using anatomical regularities according to the guideline.
Finally, a regression forest integrated with constraints of plane regularities is
applied for refining each plane. The proposed method was evaluated on a 3-D
cardiac ultrasound dataset and a synthetic dataset. Compared with other plane
extraction methods, it demonstrated an improved accuracy with a significantly
faster running time of 0.8s/volume. Furthermore, it showed the proposed
method was robust for a range abnormalities and image qualities, which would be
seen in clinical practice [22].

\textbf{Prostate}\newline
i) In 1992, Prater and Richard introduced a method for segmenting transrectal ultrasound images of the prostate using feedforward neural networks. They proposed three neural net architectures in this method. Each of these networks was trained using a small portion of a training image segmented by an expert sonographer. Although the performance of the relatively simple networks described here demonstrates the potential of neural networks for automatically segmenting ultrasound images of the prostate, there are some limitations.
One of the limitations of this method is that the networks are all variations on one theme, using data in the same row and column as the center pixel to the center pixel to obtain a mix of global and local input information, whereas many other input schemes are possible. Moreover, a wider range of them should be investigated to greater accuracy and reduced computational complexity. Another limitation of this research is that training data was taken from a maximum of three images, whereas prostate size and shape vary from individual to individual [23].
\textbf{Blood Vessels}\newline
i) In 2016, Smistad and his team presented a new method for detecting blood vessels in B-mode ultrasound images using a deep convolutional neural network. They proposed to use an ellipse fitting at each pixel in the image using a graphic processing unit (GPU) to find vessel candidate regions which are passed on to a deep neural network classifier which determines if the region contains a vessel or not. The first step finds vessel candidates and creates sub-images for each. The sub-images are then passed on to a deep neural network which identifies the sub-images belonging to vessels, and discards those that are not of vessels. The proposed method is able to determine the position and size of the vessels in images in real-time. The vessel model used in the proposed method assumes that the vessels are elliptical, while this often holds true for arteries, it may not be ideal for veins which often have a more irregular shape. Thus, the proposed method is more suited for arteries than veins [24].
\textbf{Lymph Nodes}\newline
i) Quantitative analysis of lymph nodes size, shape, morphology and
their relations in ultrasound images gives useful and reliable
information for clinical diagnosis, cancer staging, patient prognosis,
and treatment planning. It also helps obtain a better understanding
of what are solid and effective features for diagnosing lymph node
related diseases. An automatic method for segmenting lymph nodes
in ultrasound images lays a foundation for such quantitative analysis
and disease studies.
this paper proposed a novel coarse-to-fine stacked FCN (CFS-FCN)
deep learning model for lymph node segmentation in ultrasound
images. The new models learn segmentation knowledge in a coarse-
to-fine and simple-to-complex manner.

CFS-FCN model consists of multiple stages of FCNs(fully
convolutional networks). FCN A is trained to learn
segmentation(filtering) knowledge from the raw input image to
produce a segmentation label map (intermediate result) that shows
all the areas visually similar to lymph nodes (non-expert knowledge,
possibly with false positives). Then FCN B is trained to use the
intermediate result combined with the raw image to produce the final
(real) lymph node segmentation.
Our model can be extended to have more than 2 stages. The number
of stages used depends on different applications. There are three
main design issues to this model: (i) The architecture of FCN
modules (FCN A and FCN B); (ii) the coarse-to-fine label maps (the
intermediate label map and final label map); (iii) a suitable training
strategy for training CFS-FCN. The following Sub sections B, C, and
explore these three issues, respectively. In Sub section E, a post-
processing is applied to further refine the lymph node boundaries
based on the segmentation results produced by CFS-FCN
Comparing to the state-of-the-art deep learning methods for
biomedical segmentation, this method yields better segmentation of
lymph nodes in ultrasound images. they further applied a convex-
shape constraint based boundary refinement method to enhance the
quality of the segmented lymph nodes.
With accurately detected and segmented lymph nodes, quantitative
measurement and analysis can be developed for computer-aided
diagnosis and studies of lymph node related diseases using
ultrasound images [25]

\textbf{Generic Techniques}\newline
i) In case of ultrasound image segmentation, the main challenge is to deal with images is the poor quality of images, which are also affected by speckle noise. Watershed segmentation is one of the most effective methods in complex segmentation problems. The algorithm uses watershed transform applied to images to obtain the segmented regions. However, segmentation of noisy ultrasound image using watershed transform always leads to over-segmentation. So, in 2009, Bhushan introduced some techniques in order to remove the over-segmentation. The images were first contrast enhanced using histogram equalization and then median filtered using 7x7 window. Then the gradient map is generated using multi-scale morphological gradient. Then the small local minima removed by reconstruction by erosion [26].
ii) In 2018, Li and his team proposed Ultrasound image analysis and recognition techniques for improving workflow in diagnosis and treatment. Here, fully automatic techniques for applications of cardiac plane extraction, foetal weight measurement and ultrasound-CT image registration for liver surgery navigation were introduced. For standard plane extraction in 3D cardiac ultrasound, multiple cardiac landmarks defined in ultrasound cardiac examination guidelines were detected. For automatic foetal weight measurement, bi-parietal diameter (BPD), femur length (FL) and abdominal circumference (AC) are estimated by segmenting corresponding organs and regions from foetal ultrasound images. And machine learning method was used to extract ultrasound-CT image pair and later they were used for image registration. However, the accuracy can be affected by poor quality of ultrasound image [27].

\begin{figure}[htbp]
\centerline{\includegraphics[width=3in]{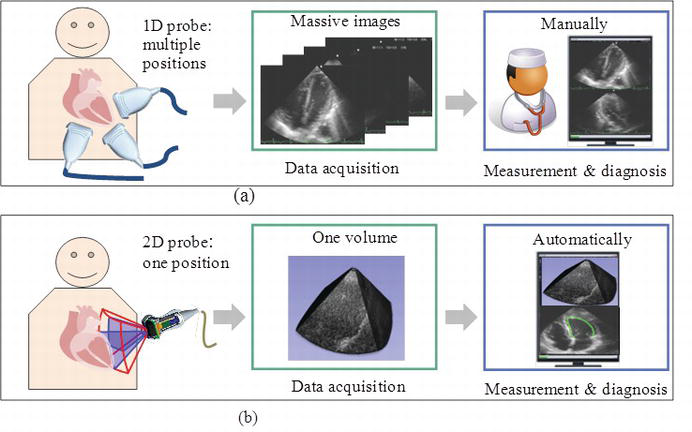}}
\caption{Figure 3.(a) Conventional 2D ultrasound diagnosis and (b) 3D ultrasound diagnosis in the future. [27]}
\label{fig}
\end{figure}

iii) A list of methods were reviewed by Wen et al., to find out the best methods in image signal processing. Methods such as Statistics Transfer, Content Aware Quantization, Image Reshaping, Data-Driver Image Restoration, and Compressed Sensing Magnetic Resonance Imaging were evaluated.  Imaging technologies such as High Dynamic Range Imaging (HDR) and compressed sensing were also reviewed. It was shown that older more traditionally used methods do not have good results in applications related to HDR and MRI, whereas, smart approaches as those mentioned above have better efficiency and perform better than heuristic approaches. Especially Content-Aware Image Retargeting and Compressed Sensing MRI are seen to have promising results [28].
 
iv) Eusebio et al., worked on semi-supervised adversarial image-to-image translation, in which, while keeping the properties of an image consistent, an image is converted to and from different domains. This could be useful in the segmentation of US images, so as not to stay limited in only one domain. Datasets like SVHN and MNIST containing handwritten letters and digits were used to perform semi-supervised translation. Semi-supervision is shown to be a better option and a good trade-off compared to its fully-supervised and unsupervised counterparts. However, the paper states that improvements in data representation is required to work better with this method [29].
 
v) Real-time runway detection using synthetic vision for infrared aerial images was worked upon by Liu et al., for which three-thresholding segmentation and Region of Interest (ROI) were used successively to segment the runway. Existing methods were compared with those used in this paper. The method is shown to perform better time-wise compared to other existing methods as well as an improved accuracy derived from the initial contour taken. Detection and delineation between runways and touch zones is done efficiently [30].

vi) Furthermore, for classification of gestures made by hand, Time Delay Neural Network, Dynamic Time Warping, Finite State Machine, Hidden Markov Model (HMM) and the common methods Convolutional Neural Network (CNN) are implemented, and were thus reviewed by Wang et al,. Considering the recognition of static hand gesture, only one or a set of few shape features are considered and tracked instead of tracking the hand motion. Highest average accuracy of 99.4% was shown by Support Vector Machine (SVM) with only the limitation dependent on the quality of the camera used [31].
 
vii) Based on EREL region detection technique and Convolutional Neural Networks (CNN), a new method called EREL-Net was proposed by Patel et al., with the objective to detect defects in bottles. One of the key features of this paper was the sparse dataset, which still supposedly gives good classification with EREL, the regions being extracted from Edge-Amplified (EA) images. State-of-the-art EREL region detectors are used to get Regions of Interest (ROI), with the further classification by CNN. The overall accuracy varied between 88-100% with a false positive rate highest at 14% only. This can therefore be considered a prospective candidate for classification of shapes in US images [32].
 
viii) Preservation and enhancement of details in a low resolution image can be done by converting it to a high dynamic range (HDR) image. The ambiguity and loss of features in US images can thus be remedied in some form by this method. Cheng et al., had their work related to the artefacts caused by quantization when going from SDR to HDR. Addition of noise with Gaussian noise generation was performed to get better results than traditional methods when it comes to computing and memory I/O limited systems [33]. 

ix) Another paper published by Subhayan et al., details the attempt to segment brain hemorrhages using MR (Magnetic Resonance) Images. For this they used Otsu's method to first to get rid of the noisy background. Next, for ventricle detection, MSER (Maximally Stable Extremal Regions) algorithm was used. Parts not being ventricles were eliminated by considering the physiology of the brain such that the regions farther from the boundaries are likely to be ventricles. Contour detection for white matter and K-means clustering is used to find the small and large objects, of which, smaller ones are more likely to be hemorrhages. The paper claims to achieve upto 50-100% sensitivity and 99% specificity. The issues faced in this project, such as false positives and noise removal are also present in US image segmentation, and thus these approaches can be considered [34].

x) MR images were also worked upon by Soltaninejad et al., who aimed at detection of Parkinson's Disease (PD). In order to classify PD, features such as ventricles, cerebrospinal fluid and others were considered with Logistic Regression, Support Vector Machines and Random Forest. Of these, Random Forest classifier was shown to have a high train and test accuracy when compared with the methods used in the same experiment, but a relatively low accuracy otherwise of 75-79%. Regardless of this, it can be said that when it comes to noisy data that can have multiple false positives, Random Forest is a better choice compared to Logistic Regression and Support Vector Machines [35].
 
xi) It develops a novel approach that provides effective and robust segmentation of color images.
By incorporating the advantages of the mean shift (MS) segmentation and the normalized cut
(Ncut) partitioning methods, the proposed method require slow computational complexity and is
therefore very feasible for real-time image segmentation processing. It preprocesses an image by
using the MS algorithm to form segmented regions that preserve the desirable discontinuity
characteristics of the image. The segmented regions are then represented by using the graph
structures, and the N-cut method is applied to perform globally optimized clustering. Because the
number of the segmented regions is much smaller than that of the image pixels, the proposed method
allows a low-dimensional image clustering with significant reduction of the complexity compared to
conventional graph-partitioning methods that are directly applied to the image pixels. In addition,
the image clustering using the segmented regions, instead of thei mage pixels, also reduces the
sensitivity to noise and results in enhanced image segmentation performance. Furthermore, to avoid
some inappropriate partitioning when considering every region as only one graph node, we develop an
improved segmentation strategy using multiple child nodes for each region. The superiority of the
the proposed method is examined and demonstrated through a large number of experiments using color
natural scene images [36].

xii) There is large consent that successful training of deep net-works requires many
thousand annotated training samples. In this pa-per, we present a network and
training strategy that relies on the strong use of data augmentation to use the
available annotated samples more efficiently. The architecture consists of a
contracting path to capture context and a symmetric expanding path that enables
precise localization. We show that such a network can be trained end-to-end from
very few images and outperforms the prior best method (a sliding-window
convolutional network) on the ISBI challenge for segmentation of neuronal structures
in electron microscopic stacks [37].

\section{Conclusion}
\label{conclusion}

The applications and possibilities that can be achieved through ultrasound imaging are seemingly endless, especially when it comes to diagnostic medicine. As reviewed above however, automatic detection is necessary over the subjective methods that are practiced today relying on radiologists and other experts. Multiple deep learning techniques are compared with each other and are even combined in some papers. With the amount of ambiguity that is prevalent in the segmentation of ultrasound images, a robust and efficient system is required. A good system that is able to accurately segment, identify and detect regions in Ultrasound images will prove to be a huge step in precise 3D reconstructions that will in turn advance the field of diagnostic medicine by multifolds. Our proposed research aims to compare deep learning techniques such as the Dual Path Network with machine learning techniques that might prove more fruitful by eliminating the blackbox feature of deep learning methods.

\section{References}
\bibliographystyle{IEEEbib}
\bibliography{egbib}

%you should add references to a .bib file (egbib.bib is an example of how a bibliography file looks like). to refer to the references in the text you should use "~\cite{Martin2001}"
\justify
[1] Adler, Robin et al. “An Ultrasonic Remote Control for Home Receivers.” (1957).\newline
[2] Haar, Gail ter. “Therapeutic applications of ultrasound.” (2007).\newline
[3] Brady, Adrian P.. “Error and discrepancy in radiology: inevitable or avoidable?” Insights into Imaging (2016).\newline
[4] Lele, Padmakar P.. “Application of ultrasound in medicine.” The New England journal of medicine 286 24 (1972): 1317-8.\newline
[5] Burckhardt, Christoph Benedikt. “Speckle in ultrasound B-mode scans.” IEEE Transactions on Sonics and Ultrasonics 25 (1978): 1-6.\newline
[6] Wagner, R. G. et al. “Statistics of Speckle in Ultrasound B-Scans.” IEEE Transactions on Sonics and Ultrasonics 30 (1983): 156-163.\newline
[7] Noble, J. Alison and Djamal Boukerroui. “Ultrasound image segmentation: a survey.” IEEE Transactions on Medical Imaging 25 (2006): 987-1010.\newline
[8] Drukker, Karen, et al. "Computerized lesion detection on breast ultrasound." Medical physics 29.7 (2002): 1438-1446.\newline
[9] K. Horsch, M. L. Giger, L. A. Venta, C. J. Vyborny, "Computerized diagnosis of breast lesions on ultrasound", Med. Phys., vol. 29, no. 2, pp. 157-164, Feb. 2002.\newline
[10] Zhou, Zhuhuang et al. “Semi-automatic breast ultrasound image segmentation based on mean shift and graph cuts.” Ultrasonic imaging 36 4 (2014): 256-76.\newline
[11] Jiang, Peng et al. “Learning-based automatic breast tumor detection and segmentation in ultrasound images.” ISBI (2012).\newline
[12] Prevost, Raphael, et al. "Kidney detection and real-time segmentation in 3D contrast-enhanced ultrasound images." 2012 9th IEEE International Symposium on Biomedical Imaging (ISBI). IEEE, 2012.\newline
[13] Marsousi, Mahdi et al. “Kidney Detection in 3-D Ultrasound Imagery via Shape-to-Volume Registration Based on Spatially Aligned Neural Network.” IEEE Journal of Biomedical and Health Informatics 23 (2019): 227-242.\newline
[14] Xie, Jun et al. “Segmentation of kidney from ultrasound images based on texture and shape priors.” IEEE Transactions on Medical Imaging 24 (2005): 45-57.\newline
[15] Raynaud, Caroline, et al, "Multi-organ detection in 3D fetal ultrasound with machine learning." Fetal, Infant and Ophthalmic Medical Image Analysis. Springer, Cham, 2017. 62-72.\newline
[16] Gomez, Alberto, et al. "Fast registration of 3D fetal ultrasound images using learned corresponding salient points." Fetal, Infant and Ophthalmic Medical Image Analysis. Springer, Cham, 2017. 33-41.\newline
[17] Sridar, Pradeeba \& Kumar, Ashnil \& Quinton, Ann \& Krishna kumar, Ramarathnam \& Feng, David Dagan Feng \& Nanan, Ralph \& Kim, Jinman "Automatic Identification of Multiple Planes of a Fetal Organ from 2D Ultrasound Images." (2016).\newline
[18] Li, Yuying and Mehdi Faraji. “EREL Selection Using Morphological Relation.” ArXiv abs/1806.03580 (2018): n. pag.
[19] Degel, Markus A. et al. “Domain and Geometry Agnostic CNNs for Left Atrium Segmentation in 3D Ultrasound.” MICCAI (2018).
[20] Yang, Ji et al. “IVUS-Net: An Intravascular Ultrasound Segmentation Network.” ICSM (2018).\newline
[21] Yang, J. C. et al. “Robust segmentation of arterial walls in intravascular ultrasound images using Dual Path U-Net.” Ultrasonics 96 (2019): 24-33.\newline
[22] Zhu, Peifei and Zisheng Li. “Guideline-based learning for standard plane extraction in 3-D echocardiography.” Journal of medical imaging 5 4 (2018): 044503.\newline
[23] Prater, James S., and William D. Richard. "Segmenting ultrasound images of the prostate using neural networks." Ultrasonic Imaging 14.2 (1992): 159-185\newline
[24] Smistad, Erik, and Lasse Løvstakken. "Vessel detection in ultrasound images using deep convolutional neural networks." Deep Learning and Data Labeling for Medical Applications. Springer, Cham, 2016. 30-38.\newline
[25] Zhang, Yizhe et al. “Coarse-to-Fine Stacked Fully Convolutional Nets for lymph node segmentation in ultrasound images.” 2016 IEEE International Conference on Bioinformatics and Biomedicine (BIBM) (2016): 443-448.\newline
[26] Bhushan, Chitresh. “Ultrasound Image Segmentation.” (2009).
[27] Li, Zisheng et al, “Automatic Image Analysis and Recognition for Ultrasound Diagnosis and Treatment in Cardiac, Obstetrics and Radiology.” (2018).\newline
[28] Hajari, Nasim et al. “Spatio-Temporal Eye Gaze Data Analysis to Better Understand Team Cognition.” ICSM (2018).\newline
[29] Eusebio, Jose et al. “Semi-supervised Adversarial Image-to-Image Translation.” ICSM (2018).\newline
[30] Liu, Changjiang et al. “Synthetic Vision Assisted Real-Time Runway Detection for Infrared Aerial Images.” ICSM (2018).\newline
[31] Wang, Taiqian et al. “A Survey on Vision-Based Hand Gesture Recognition.” ICSM (2018).\newline
[32] Patel, Nikunjkumar et al. “EREL-Net: A Remedy for Industrial Bottle Defect Detection.” ICSM (2018).\newline
[33] Mukherjee, Subhayan et al. “Adaptive Dithering Using Curved Markov-Gaussian Noise in the Quantized Domain for Mapping SDR to HDR Image.” ICSM (2018).\newline
[34] Mukherjee, Subhayan et al. “Atlas-Free Method of Periventricular Hemorrhage Detection from Preterm Infants' T1 MR Images.” ICSM (2018).\newline
[35] Soltaninejad, Sara et al. “Towards the Identification of Parkinson's Disease Using only T1 MR Images.” ArXiv abs/1806.07489 (2018): n. pag.\newline
[36] Tao, Wenbing et al. “Color Image Segmentation Based on Mean Shift and Normalized Cuts.” IEEE Transactions on Systems, Man, and Cybernetics, Part B (Cybernetics) 37 (2007): 1382-1389.\newline
[37] Ronneberger, Olaf et al. “U-Net: Convolutional Networks for Biomedical Image Segmentation.” ArXiv abs/1505.04597 (2015): n. pag. \newline

\end{document}